\documentclass[pra]{revtex4}
\usepackage{amssymb}
\usepackage{amsmath}
\usepackage{graphicx}
\usepackage{dcolumn}
\usepackage[center]{subfigure}
\usepackage{array}
\usepackage{booktabs}
\usepackage{multirow}
\usepackage{float}
\usepackage{color}
\usepackage{ulem}

\begin{document}

\title{Vortex gap solitons in spin-orbit-coupled Bose-Einstein condensates
with competing nonlinearities}
\author{Xiaoxi Xu$^{1}$}
\author{Feiyan Zhao$^{1}$}
\author{Yangui Zhou$^{1}$}
\author{Bin Liu$^{1}$}
\author{Xunda Jiang$^{1}$}
\email{jxd194911@163.com}
\author{Boris A. Malomed$^{2,3}$}
\author{Yongyao Li$^{1,4}$}
\affiliation{$^{1}$ School of Physics and Optoelectronic Engineering, Foshan University,
Foshan 528000, China \\
$^{2}$ Department of Physical Electronics, School of Electrical Engineering,
Faculty of Engineering, Tel Aviv University, Tel Aviv 69978, Israel \\
$^{3}$ Instituto de Alta Investigaci\'{o}n, Universidad de Tarapac\'{a},
Casilla 7D, Arica, Chile \\
$^{4}$ Guangdong-Hong Kong-Macao Joint Laboratory for Intelligent Micro-Nano
Optoelectronic Technology, Foshan University, Foshan 528000, China}
\date{\today}

\begin{abstract}
The formation and dynamics of \textit{full vortex gap solitons} (FVGSs) is
investigated in two-component Bose-Einstein condensates with spin-orbit
coupling (SOC), Zeeman splitting (ZS), and competing cubic and quintic
nonlinear terms, while the usual kinetic energy is neglected, assuming that
it is much smaller than the SOC and ZS terms. Unlike previous SOC system
with the cubic-only attractive nonlinearity, in which solely \textit{%
semi-vortices} may be stable, with the vorticity carried by a single
component, the present system supports stable FVGS states, with the
vorticity present in both components (such states are called here
\textquotedblleft full vortex solitons", to stress the difference
from the half-vortices). They populate the bandgap in the system's linear
spectrum. In the case of the cubic self-attraction and quintic repulsion,
stable FVGSs with a positive effective mass exist near the top of the
bandgap. On the contrary, the system with cubic self-repulsion and quintic
attraction produces stable FVGSs with a negative mass near the bottom of the
bandgap. Mobility and collisions of FVGSs with different topological charges
are investigated too.\newline
\textbf{Key words}: Full vortex gap solitons, cubic-quintic nonlinearity,
soliton dynamics
\end{abstract}

\maketitle

\section{Introduction}

Since the Bose-Einstein condensates (BECs) have been created in ultracold
gases of alkali metals \cite{Anderson1995,Bradley1995,Davis1995}, they have
become a universal platform to study many phenomena in superfluids, such as
vortices \cite{Williams1999,Kasamatsu2003}, solitons \cite%
{Burger1999,Anderson2001} and many other dynamical effects \cite%
{Heinzen2000,Morsch2006,Eckardt2009}. In their straightforward form,
one-dimensional (1D) solitons stably exist in BECs without external traps,
while two- and three-dimensional (2D and 3D) solitons are made unstable by
the presence of the collapse, driven by the cubic self-attraction \cite%
{Berg1998,Sulem1999,Fibich2015,Malomed2005,Malomed2016}.

A significant direction in the work with atomic BEC is their use for
emulation of various phenomena which are known in a much more complex form
in condensed-matter physics \cite{emulator,Ahufinger}. In particular,
synthetic spin-orbit coupling (SOC), which simulates this fundamental effect
well known in physics of semiconductors \cite{Dresselhaus,Bychkov}, has been
realized in two-component BEC\ by using counterpropagating Raman laser beams
to engineer the necessary coupling between the two atomic states in the
binary condensate \cite{Lin2011,Liu2014,Wu2016}. SOC is a source of many
remarkable findings in BECs, including various extended patterns, vortices,
and solitons \cite%
{Wang2010,Kawakami2011,Kawakami2012,Conduit2012,Liu2012,Zhou2013,Zezyulin2013,Sakaguchi2013,Xu2013,Kartashov2013,Kartashov2017}%
. A noteworthy fact is that SOC affects ground-state properties and enhances
stability of the system, suppressing the critical collapse in the
two-dimensional BEC with the self- and cross-attractive cubic nonlinearity,
thus making it possible to predict absolutely stable 2D solitons \cite%
{Sakaguchi2014,Sakaguchi2016}. In 3D, full suppression of the respective
supercritical collapse is not possible, but SOC was predicted to maintain
metastable 3D solitons \cite{Zhang2015}. The 2D and 3D solitons created by
SOC may be \textit{semi-vortices} (SVs), built of zero-vorticity and
vortical components, or \textit{mixed modes} (MMs), which combine
zero-vorticity and vortex terms in both components \cite%
{Sakaguchi2014,Zhang2015,Sakaguchi2016,Malomed2018}. In the context of
binary dipolar BECs, SOC may support stable anisotropic 2D composite
solitons \cite{Li2017}.\

Under the action of strong Zeeman splitting (ZS) added to the SOC system,
the MM states are transformed into stable SV gap solitons \cite%
{Sakaguchi2018}. The same effect stabilizes 2D solitons, supported by
long-range dipole-dipole interactions between atoms carrying permanent
magnetic moments, close to the edge of the ZS-induced bandgap \cite{Liao2017}%
. However, the dominant component of the 2D composite gap solitons, reported
in Refs. \cite{Liao2017} and \cite{Sakaguchi2018}, is a fundamental mode
without any phase structure. The creation of stable \textit{full-vortex} gap
solitons (FVGSs) in the SOC-affected BEC system, which, unlike the SVs,
would carry vorticity in \textit{both components}, remains a challenging
problem. In fact, such solitons were considered, in the form of
\textquotedblleft excited states" of SVs and MMs, in the system which did
not include the ZS terms \cite{Sakaguchi2014}. However, they all were found
to be completely unstable states.

To construct stable FVGSs in this work, we use competing cubic and quintic
nonlinearities. This approach is suggested by the prediction of robust 2D
\cite{Li2018,Zhang2019,Lin2021,Zheng2021} and 3D \cite{Taruell} vortex
states in \textit{quantum droplets, }which are stabilized by the
Lee-Huang-Yang effect, i.e., additional quartic self-repulsive nonlinearity,
induced by quantum fluctuations around mean-field states. The stability is
provided by the competition of the quartic self-repulsion with the cubic
mean-field attraction between components in the binary BEC \cite%
{Petrov2015,Petrov2016,Schmitt2016,Cabrera2018}. A similar possibility is to
consider higher-order nonlinearities of the mean-field origin in BEC \cite%
{Yamaguchi1979}. In that connection, it is relevant to mention that, in
addition to the common two-body (cubic) interaction, the three-body
(quintic) term can be adjusted by means of the Feshbach resonance \cite%
{Marcelli1999,Smith2014,Adhikari2017}. In previous studies, it was
demonstrated that competing cubic self-attraction and quintic repulsion
could provide the stability of vortex solitons in BECs (in the absence of
SOC) \cite{Mithun2013,Abdullaev2001,Gao2018,PhysicaD,Xu2021}.

In this work, we consider a possibility to use the interplay of the
competing nonlinearities, SOC, and ZS to support stable FVGSs in the
two-component BEC system. We produce the density distribution and phase
structure for families of stationary vortex solitons with different
topological charges, and identify their stability. Their mobility and
collisions are addressed too.

The paper is organized as follows. In Section 2, we introduce the model and
FVGS states in it. In Section 3, we study the dynamics, mobility and
collisions of the FVGSs by means of systematic simulations. The paper is
concluded by Section 4.

\section{The model and stationary solutions}

\subsection{The coupled Gross-Pitaevskii equations}

We consider the binary BEC which, in the mean-field approximation, is
modelled by the system of 2D Gross-Pitaevskii equations (GPEs) for wave
functions $\tilde{\psi}_{\pm }(\tilde{x},\tilde{y})$ of two components of
the condensate. The equations include the Rashba-type SOC \cite%
{Rashba,Sakaguchi2016} with strength $\lambda $, ZS with chemical-potential
difference $2\tilde{\Omega}$ (both $\lambda $ and $\tilde{\Omega}$ are
defined to be positive), and a combination of cubic and quintic terms, which
represents two-body and three-body interactions, respectively \cite%
{Abdullaev2001,Braaten2002,Bulgac2002,Michinel1,Salerno,Kofane}:
\begin{equation}
i\hbar {\frac{\partial }{\partial \tilde{t}}}\tilde{\psi}_{\pm }=-{\frac{\hbar ^{2}}{%
2m}}\nabla _{\pm }^{2}\tilde{\psi}\pm \lambda \left( {\frac{\partial }{%
\partial \tilde{x}}}\mp i{\frac{\partial }{\partial \tilde{y}}}\right)
\tilde{\psi}_{\mp }\mp \tilde{\Omega}\tilde{\psi}_{\pm }+g_{1}(|\tilde{\psi}%
_{\pm }|^{2}+|\tilde{\psi}_{\mp }|^{2})\tilde{\psi}_{\pm }+g_{2}(|\tilde{\psi%
}_{\pm }|^{2}+|\tilde{\psi}_{\mp }|^{2})_{\pm }^{2}\tilde{\psi}  \label{GP}
\end{equation}%
(the tilde designates variables and coefficients which are replaced below by
their scaled counterparts). The strength of the two-body interaction,
\begin{equation}
g_{1}=2\sqrt{2\pi }\hbar ^{2}a/\left( a_{\perp }m\right) ,  \label{g1}
\end{equation}%
is proportional to the \textit{s}-wave scattering length $a$ ($a>0$ and $a<0$
represent the repulsive and attractive interactions, respectively), and
inversely proportional to size $a_{\perp }$ of the confinement in the
transverse direction \cite{Dalfovo1999}. The ZS effect can be induced by an
artificial magnetic field in BECs, which breaks the symmetry between the two
components of the binary wave function and produces a bandgap in the energy
spectrum of the system \cite{Lin2011,Quay2010,Goldman2014}.

The nonlinearity in Eq. (\ref{GP}) is written in the Manakov's form \cite%
{Manakov}, i.e., with equal coefficients of the self- and
cross-interactions, which is a relevant approximation for mixtures of two
different atomic states \cite{Manakov-BEC,Manakov-BEC2}. The magnitude and
sign of $a$ can be adjusted to necessary values in Eq. (\ref{g1}) by means
of the Feshbach resonance \cite{Feshbach}. Further, the strength of the
quintic nonlinearity is
\begin{equation}
g_{2}=\frac{4\sqrt{3}\hbar ^{2}a^{4}}{ma_{\perp }^{2}}\left[ d_{1}+d_{2}\tan
\left( s_{0}\ln \frac{|a|}{|a_{0}|}+\frac{1}{2}\right) \right] ,  \label{g2}
\end{equation}%
where $a_{0}$ is the value of $a$ at which a three-body bound state appears,
$d_{1,2}$ and $s_{0}$ being coefficients given in Refs. \cite%
{Bulgac2002,Braaten2002}. It is known that $g_{2}$ can be tuned to arbitrary
positive or negative value. In this work, we focus on the competition
between the cubic and quintic nonlinearities with opposite signs, which
implies $g_{1}g_{2}<0$. As mentioned above, the \textit{s}-wave scattering
length $a$ in Eq. (\ref{g1}) can be tuned to a desired value by means of the
Feshbach-resonance technique, therefore the value of $g_{1}$ is also
adjustable.

Following Refs. \cite{Li2017} and \cite{Sakaguchi2018}, we focus on the
situation when the SOC energy is much larger than the kinetic energy, which
implies the consideration of configurations with sufficiently large
characteristic length scales, $l\gg \hbar ^{2}/\left( m\lambda \right) $. In
this case, terms $\sim \nabla ^{2}\psi _{\pm }$ in Eq. (\ref{GP}) may be
neglected. By means of scaling,%
\begin{equation}
\tilde{\psi}_{\pm }=\sqrt{\left\vert g_{1}/g_{2}\right\vert }\psi _{\pm
},\left( \tilde{x},\tilde{y}\right) =\left( \lambda \left\vert
g_{2}\right\vert /g_{1}^{2}\right) \left( x,y\right) ,\tilde{t}=\left( \hbar
\left\vert g_{2}\right\vert /g_{1}^{2}\right) t,\tilde{\Omega}=\left(
g_{1}^{2}/\left\vert g_{2}\right\vert \right) \Omega,  \label{rescaling}
\end{equation}
the so simplified GPE system is cast in the following form, in which the
nonlinearity and SOC coefficients are set to be $1$:
\begin{eqnarray}
&&i{\frac{\partial }{\partial t}}\psi _{+}=\left( {\frac{\partial }{\partial
x}}-i{\frac{\partial }{\partial y}}\right) \psi _{-}-\Omega \psi _{+}-(|\psi
_{+}|^{2}+|\psi _{-}|^{2})\psi _{+}+\left( |\psi _{+}|^{2}+|\psi
_{-}|^{2}\right) ^{2}\psi _{+},  \notag \\
&&i{\frac{\partial }{\partial t}}\psi _{-}=-\left( {\frac{\partial }{%
\partial x}}+i{\frac{\partial }{\partial y}}\right) \psi _{+}+\Omega \psi
_{-}-(|\psi _{-}|^{2}+|\psi _{+}|^{2})\psi _{-}+\left( |\psi _{-}|^{2}+|\psi
_{+}|^{2}\right) ^{2}\psi _{-},  \label{1}
\end{eqnarray}%
in the case of the competition between cubic attraction and quintic
repulsion, or
\begin{eqnarray}
&&i{\frac{\partial }{\partial t}}\psi _{+}=\left( {\frac{\partial }{\partial
x}}-i{\frac{\partial }{\partial y}}\right) \psi _{-}-\Omega \psi _{+}+(|\psi
_{+}|^{2}+|\psi _{-}|^{2})\psi _{+}-(|\psi _{+}|^{2}+|\psi
_{-}|^{2})^{2}\psi _{+},  \notag \\
&&i{\frac{\partial }{\partial t}}\psi _{-}=-\left( {\frac{\partial }{%
\partial x}}+i{\frac{\partial }{\partial y}}\right) \psi _{+}+\Omega \psi
_{-}+(|\psi _{-}|^{2}+|\psi _{+}|^{2})\psi _{-}-(|\psi _{+}|^{2}+|\psi
_{-}|^{2})^{2}\psi _{-},  \label{2}
\end{eqnarray}%
in the opposite case of the cubic repulsion and quintic attraction. The
remaining ZS parameter $\Omega $ may be also fixed by rescaling, but we keep
it in the equations, as it is relevant to display some results for a fixed
norm of the solution, while varying $\Omega $.

Actually, Eqs. (\ref{1}) and (\ref{2}) can be transformed into each other by
simple substitution
\begin{equation}
\psi _{\pm }\rightarrow \psi _{\mp }^{\ast },  \label{convert}
\end{equation}%
where $\ast $ stands for the complex conjugate; however, this transformation
does not apply to the full system of Eqs. (\ref{GP}).

\begin{figure}[th]
\centering{\includegraphics[scale=0.35]{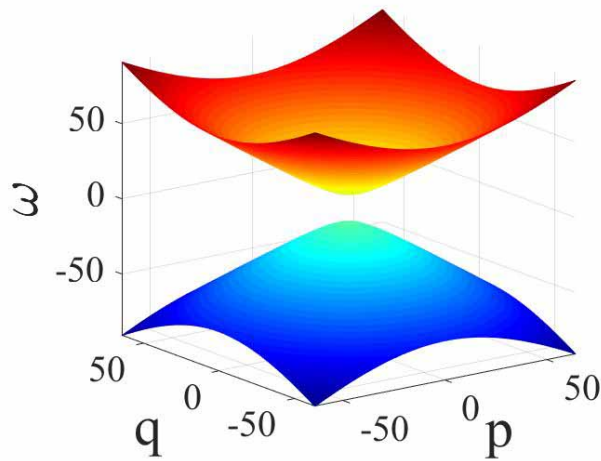}}
\caption{(Color online) Dispersion relation given by Eq. (\protect\ref%
{bandgap}) with $\Omega =10$.}
\label{Bandgapfigure}
\end{figure}

The linearization of Eqs. (\ref{1}) and (\ref{2}) is carried out by
substituting
\begin{equation}
\psi _{\pm }=\varphi _{\pm }\exp \left( -i\omega t+ipx+iqy\right)
\label{Linearization}
\end{equation}%
with constant amplitudes $\varphi _{\pm }$, which leads to the dispersion
relation between the frequency $\omega $ and wave numbers $p$ and $q$ \cite%
{Li2017}
\begin{equation}
\omega ^{2}=\Omega ^{2}+(p^{2}+q^{2}),  \label{bandgap}
\end{equation}%
which is plotted in Fig. \ref{Bandgapfigure}. Obviously, Eq. (\ref{bandgap})
with $\Omega \neq 0$ gives rise to the bandgap of width $2\Omega $, which
may be populated by gap solitons.

\subsection{Stationary gap solitons existing at bottom and top edges of the
bandgap.}

Soliton solutions of Eqs. (\ref{1}) and (\ref{2}) with chemical potential $%
\mu $ are looked for as
\begin{equation}
\psi _{\pm }=e^{-i\mu t}\phi _{\pm }(x,y),  \label{solutions}
\end{equation}%
where $\phi _{\pm }$ are stationary wave functions. At the bottom and top
edges of the bandgap, the chemical potential of the soliton is defined as
\begin{equation}
\mu =%
\begin{cases}
\Omega -\Delta \mu ,\quad & \mathrm{top}\text{ }\mathrm{of}\text{ }\mathrm{%
the}\text{ }\mathrm{bandgap,} \\
-\Omega +\Delta \mu ,\quad & \mathrm{bottom}\text{ }\mathrm{of}\text{ }%
\mathrm{the}\text{ }\mathrm{bandgap,}%
\end{cases}
\label{mu}
\end{equation}%
with
\begin{equation}
0<\Delta \mu <<\Omega .  \label{<<}
\end{equation}%
The total norm of the two-component wave function, defined by
\begin{equation}
N=N_{+}+N_{-}\equiv \int \int \left[ |\psi _{+}(x,y)|^{2}+|\psi
_{-}(x,y)|^{2}\right] dxdy,  \label{Norm}
\end{equation}%
is a dynamical invariant of the system. The value of the norm is a natural
characteristic of the solitons.

Below, we analyze the existence of the gap solitons in the edge regions
which are defined as per Eq. (\ref{mu}). The analysis is presented here for
Eq. (\ref{2}), as, near the bottom and top edges of the bandgap Eq. (\ref{mu}%
), the transformation provided by Eq. (\ref{convert}) automatically produces
solutions for the stationary wave functions $\phi _{\pm }$, defined as per
Eq. (\ref{solutions}) and Eq. (\ref{1}):
\begin{equation}
\phi _{\mp }^{\mathrm{(top(stable);bottom(unstable))}}\left\{ \mathrm{Eq.~}%
\text{(\ref{1})}\right\} =\left[ \phi _{\pm }^{\mathrm{%
(bottom(stable);top(unstable))}}\left\{ \mathrm{Eq.~}\text{(\ref{2})}%
\right\} \right] ^{\ast }.  \label{J2symmetry}
\end{equation}%
The Lagrangian of Eq. (\ref{2}) is%
\begin{equation}
\mathcal{L}=\int \int \left( \frac{i}{2}\sum_{+,-}\psi _{\pm }^{\ast }\frac{%
\partial \psi _{\pm }}{\partial t}+\mathrm{c.c.}\right) dxdy-H,  \label{L}
\end{equation}%
where $\mathrm{c.c.}$ stands for the complex-conjugate expression, and the
Hamiltonian (another dynamical invariant of the system) is%
\begin{eqnarray}
H &=&\int \int \left\{ \left[ \frac{1}{2}\left( \psi _{+}^{\ast }\frac{%
\partial \psi _{-}}{\partial x}-i\psi _{+}^{\ast }\frac{\partial \psi _{-}}{%
\partial y}\right) +\mathrm{c.c.}\right] +\Omega \left( \left\vert \psi
_{-}\right\vert ^{2}-\left\vert \psi _{+}\right\vert ^{2}\right) \right.
\notag \\
&&\left. +\frac{1}{2}\left( \left\vert \psi _{+}\right\vert ^{2}+\left\vert
\psi _{-}\right\vert ^{2}\right) ^{2}-\frac{1}{3}\left( \left\vert \psi
_{+}\right\vert ^{2}+\left\vert \psi _{-}\right\vert ^{2}\right)
^{3}\right\} dxdy.  \label{H}
\end{eqnarray}%
The application of the Noether's theorem \cite{Noether} to Eq. (\ref{L})
produces the conserved momentum in the same form as in Schr\"{o}dinger
equations, namely,%
\begin{equation}
\mathbf{P}=i\int \int [\nabla (\psi _{+}^{\ast })\psi _{+}+\nabla (\psi
_{-}^{\ast })\psi _{-}]dxdy.  \label{P}
\end{equation}

On the other hand, the angular momentum, which may be conveniently written
in terms of polar coordinates $\left( r,\theta \right) $ as%
\begin{equation}
\Lambda =i\int_{0}^{\infty }rdr\int_{0}^{2\pi }d\theta \sum_{+,-}\frac{%
\partial \psi _{\pm }^{\ast }}{\partial \theta }\psi _{\pm },  \label{Lambda}
\end{equation}%
is not conserved in the framework of the SOC system. Indeed, rewriting Eq. (%
\ref{2}) in terms of $\left( r,\theta \right) $,%
\begin{eqnarray}
&&i{\frac{\partial }{\partial t}}\psi _{+}=e^{-i\theta }\left( {\frac{%
\partial }{\partial r}}-\frac{i}{r}{\frac{\partial }{\partial \theta }}%
\right) \psi _{-}-\Omega \psi _{+}+(|\psi _{+}|^{2}+|\psi _{-}|^{2})\psi
_{+}-(|\psi _{+}|^{2}+|\psi _{-}|^{2})^{2}\psi _{+},  \notag \\
&&i{\frac{\partial }{\partial t}}\psi _{-}=-e^{i\theta }\left( {\frac{%
\partial }{\partial r}}+\frac{i}{r}{\frac{\partial }{\partial \theta }}%
\right) \psi _{+}+\Omega \psi _{-}+(|\psi _{-}|^{2}+|\psi _{+}|^{2})\psi
_{-}-(|\psi _{+}|^{2}+|\psi _{-}|^{2})^{2}\psi _{-},  \label{2'}
\end{eqnarray}%
one obtains%
\begin{equation}
\frac{d\Lambda }{dt}=2\int_{0}^{\infty }rdr\int_{0}^{2\pi }d\theta ~\mathrm{%
Im}\left( e^{i\theta }\psi _{+}\frac{\partial \psi _{-}^{\ast }}{\partial r}%
\right) .  \label{d/dt}
\end{equation}

Substituting Eq. (\ref{solutions}) into Eq. (\ref{2}) yields the following
equations:
\begin{eqnarray}
&&\mu \phi _{+}=\left( {\frac{\partial }{\partial x}}-i{\frac{\partial }{%
\partial y}}\right) \phi _{-}-\Omega \phi _{+}+(|\phi _{+}|^{2}+|\phi
_{-}|^{2})\phi _{+}-\left( |\phi _{+}|^{2}+|\phi _{-}|^{2}\right) ^{2}\phi
_{+},  \notag \\
&&\mu \phi _{-}=-\left( {\frac{\partial }{\partial x}}+i{\frac{\partial }{%
\partial y}}\right) \phi _{+}+\Omega \phi _{-}+(|\phi _{-}|^{2}+|\phi
_{+}|^{2})\phi _{-}-\left( |\phi _{-}|^{2}+|\phi _{+}|^{2}\right) ^{2}\phi
_{-}.  \label{miu1}
\end{eqnarray}%
In particular, near the bottom edge of the bandgap, which is defined as per
the bottom line of Eq. (\ref{mu}), the first equation of Eq. (\ref{miu1})
amounts, in the lowest approximation, to a linear relation (cf. Ref. \cite%
{Sakaguchi2016}),
\begin{equation}
\phi _{-}\approx \left( 2\Omega \right) ^{-1}\left( {\frac{\partial }{%
\partial x}}+i{\frac{\partial }{\partial y}}\right) \phi _{+}.  \label{phi-}
\end{equation}%
Substituting this in the first equation of Eq. (\ref{miu1}), one obtains
\begin{equation}
-\Delta \mu \cdot \phi _{+}=-\left( 2\Omega \right) ^{-1}\left( {\frac{%
\partial ^{2}}{\partial x^{2}}}+{\frac{\partial ^{2}}{\partial y^{2}}}%
\right) \phi _{+}-|\phi _{+}|^{2}\phi _{+}+|\phi _{+}|^{4}\phi _{+}.
\label{stablephi+}
\end{equation}%
This equation, which combines the effective cubic self-attraction and
quintic repulsion, has a chance to produce stable 2D solitons \cite%
{Michinel2,Berntson}. It is well known that such solitons, with all values
of the embedded vorticity (winding number), exist in interval \cite%
{Bulgaria,Canada,Berntson}
\begin{equation}
0<\Delta \mu <3/16,  \label{3/16}
\end{equation}%
with the respective values of the norm
\begin{equation}
N(\Delta \mu \rightarrow 0)\rightarrow 0,N\left( \Delta \mu \rightarrow
3/16\right) \rightarrow \infty .  \label{Nlimits}
\end{equation}

Near the top edge of the bandgap, defined as per the top line of Eq. (\ref%
{mu}), approximate relation (\ref{phi-}) is replaced by one following from
the second equation of Eq. (\ref{miu1}):
\begin{equation}
\phi _{+}\approx {\frac{1}{2\Omega }}\left( {\frac{\partial }{\partial x}}-i{%
\frac{\partial }{\partial y}}\right) \phi _{-}.  \label{phi+}
\end{equation}%
Substituting it into the second equation of Eq. (\ref{miu1}), one obtains
\begin{equation}
-\Delta \mu \cdot \phi _{-}=-{\frac{1}{2\Omega }}\left( {\frac{\partial ^{2}%
}{\partial x^{2}}}+{\frac{\partial ^{2}}{\partial y^{2}}}\right) \phi
_{-}+|\phi _{-}|^{2}\phi _{-}-|\phi _{-}|^{4}\phi _{-},  \label{unstablephi-}
\end{equation}%
cf. Eq. (\ref{stablephi+}). It is obvious that this equation, which
combines the nonlinear terms with the signs opposite to those in Eq. (\ref%
{stablephi+}), i.e., cubic self-repulsion and quintic attraction, may
produce only unstable 2D solitons, which are subject to the destructive
effect of the supercritical collapse.

\begin{figure}[h]
\centering{\ \includegraphics[scale=0.7]{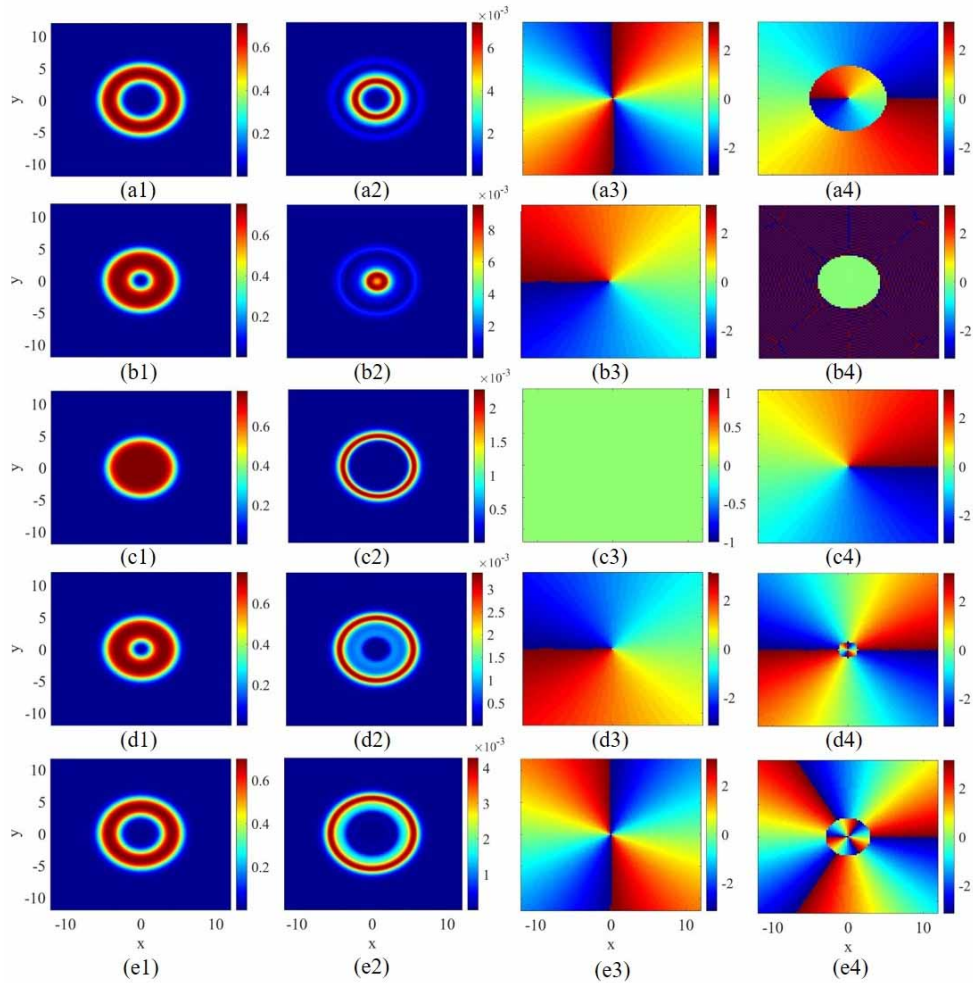}}
\caption{(Color online) Density patterns, $|\protect\phi _{+}\left(
x,y)\right) |^{2}$ and $|\protect\phi _{-}\left( x,y\right) |^{2}$ (the
first and second columns, respectively), and phase distributions in $\protect%
\phi _{+}\left( x,y\right) $ and $\protect\phi _{-}\left( x,y\right) $ (the
third and fourth columns, respectively) for the two components of stable
solitons located near the bottom edge of the bandgap, see Eq. (\protect\ref%
{mu}), as produced by the numerical solution of Eq. (\protect\ref{miu1}).
Values of the winding numbers of the components, from top to bottom, are: $%
(S_{+},S_{-})=(-2,-1)$, $(-1,0)$, $(0,1)$, $(1,2)$, and $(2,3)$. In all the
panels, we fix the total norm as $N=50$ (see Eq. (\protect\ref{Norm}), and
the ZS strength is $\Omega =5$.}
\label{bottomdensity}
\end{figure}

\section{Numerical results}

\subsection{Stationary solutions for the vortex gap solitons}

The analysis presented in the previous section indicates that stable
stationary solutions for gap solitons may exist near the bottom edge of the
bandgap produced by Eq. (\ref{2}) or, according to Eqs. (\ref{convert}) and (%
\ref{J2symmetry}), near the top edge in the case of Eq. (\ref{1}). The
initial guess for solutions of stationary Eq. (\ref{miu1}) was
obtained by means of the imaginary-time method, applied to the
time-dependent version of Eq. (\ref{stablephi+}). Then, iterating the
initial solution in the framework of the power-conserved squared-operator
method \cite{Yang}, stationary solutions were produced in the numerically
accurate form. Lastly, stability of the soliton solutions was tested by
real-time simulations of Eq. (\ref{2}), running from $t=0$ to $1000$. Actually, this
is a very long time, which corresponds, roughly, to $\sim 1000$ solitons'
relaxation time, thus making conclusions about the stability of the
stationary states fully reliable.

Note that the topological charges (winding numbers) of two components of the
stationary solitons are subject to a simple relation \cite{Sakaguchi2014}
[the same relation follows from Eq. (\ref{phi+})],
\begin{equation}
S_{+}=S_{-}-1.  \label{S+S-}
\end{equation}

Typical examples of the numerically constructed stable stationary solution
for the solitons, located close to the bottom edge of the bandgap, with $%
(S_{+},S_{-})=(-2,-1),(-1,0),(0,1),(1,2)$ and $(2,3)$, are displayed in Fig. %
\ref{bottomdensity}. In particular, all the two-component modes in which
both $S_{\pm }$ are different from zero represent FVGSs. Thus, these
composite modes may indeed be stable, unlike their completely unstable
counterparts in the form of \textquotedblleft excited states", in the usual
SOC system with the cubic self-attraction \cite{Sakaguchi2014}. Further,
Fig. \ref{bottomdensity} shows that component $\phi _{+}$ gives a dominant
contribution to norm Eq.(\ref{Norm}), while $\phi _{-}$ plays a subordinate
role (i.e., $N_{-}\ll N_{+}$, ), being mainly distributed along the inner
rim of $\phi _{+}$ in the case of $|S_{+}|>|S_{-}|$, or along the outer rim
in the opposite case, $|S_{+}|<|S_{-}|$. For the FVGS modes located near the
top edge of the bandgap, Eq. (\ref{phi+}) indicates that $\phi _{-}$ plays a
dominant role, while $\phi _{+}$ is a subordinate component $N_{-}\gg N_{+}$%
, which is mainly distributed along the inner or outer rim of $\phi _{-}$
for $|S_{+}|<|S_{-}|$ or $|S_{+}|>|S_{-}|$, respectively.

\begin{figure}[tp]
\centering{\ \includegraphics[scale=0.65]{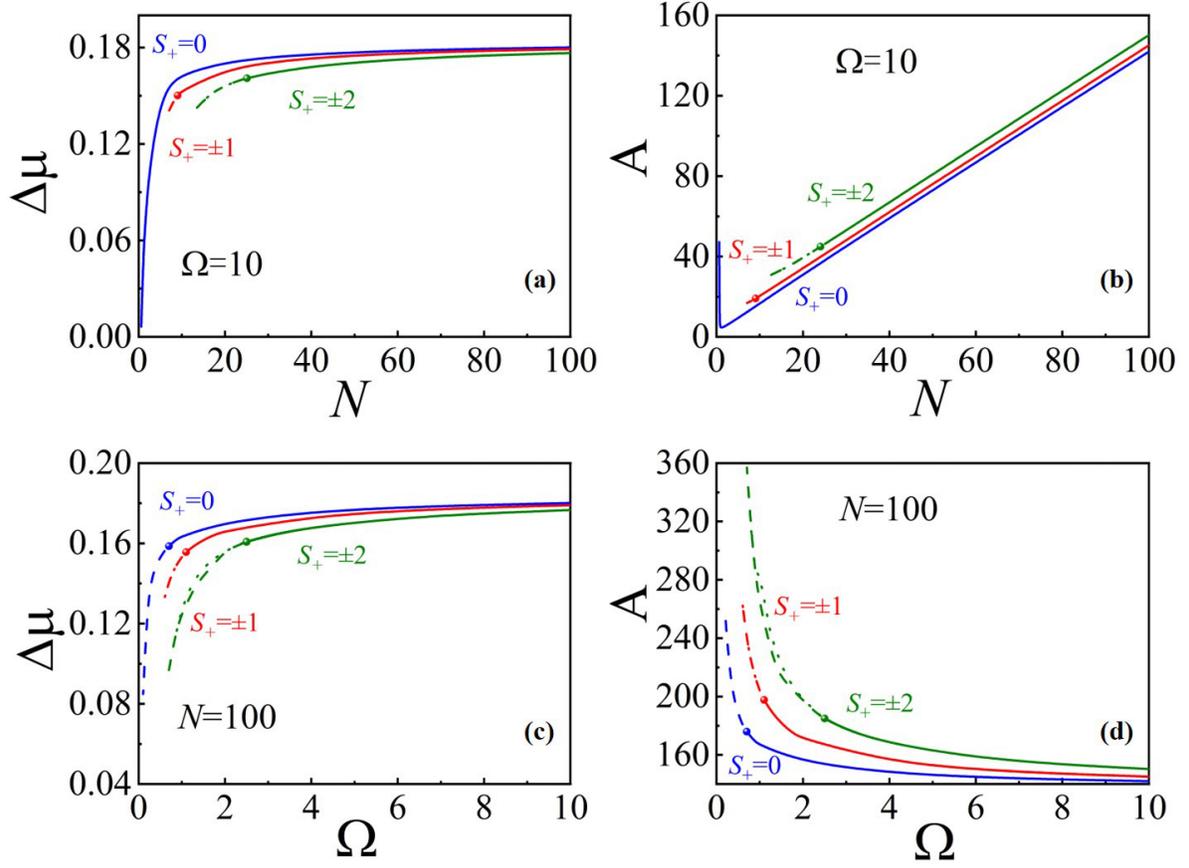}}
\caption{(Color online) The chemical-potential shift, defined as per Eq. (%
\protect\ref{mu}), and the effective area of the dominant component, defined
as per Eq. (\protect\ref{A}), versus the total norm $N$, at a fixed value of
the ZS strength $\Omega $ (a,b), and versus $\Omega $ at a fixed value of $N$
(c,d). The results are produced by the numerical solution of Eq. (\protect
\ref{miu1}). Blue, red and green curves represent the dependences for the
soliton families with the vorticity in the dominant component $S_{+}=0$, $%
\pm 1$ and $\pm 2$, respectively, while the vorticity in the other component
is given by Eq. (\protect\ref{S+S-}). Solid curves represent stable
branches, which correspond to $N>N^{\mathrm{c2}}$ and $\Omega >\Omega ^{%
\mathrm{c2}}$ (see the text). These stability boundaries are designated by
color balls on the curves. Dashed and dotted curves represent unstable
branches, at $N<N^{\mathrm{c2}}$ and $\Omega <\Omega ^{\mathrm{c2}}$ (the
dashed curves pertain to $S_{+}=0$, $-1$ and $-2$, and the dotted ones
correspond to $S_{+}=+1$ and $+2$. The stability of identified by means of
systematic simulations of Eq. (\protect\ref{2}).}
\label{properties}
\end{figure}

Characteristics of these soliton families are summarized in Fig. \ref%
{properties}, in which we plot the shift of the chemical potential, $\Delta
\mu $, defined as per the bottom line of Eq. (\ref{mu}), and the effective
area of the soliton's dominant component ($\phi _{\mathrm{d}}$),%
\begin{equation}
A\equiv \frac{\left( \int \int |\phi _{\mathrm{d}}\left( x,y\right)
|^{2}dxdy\right) ^{2}}{\int \int |\phi _{\mathrm{d}}\left( x,y\right)
|^{4}dxdy},  \label{A}
\end{equation}%
versus the total norm $N$ and ZS strength $\Omega $. Note that $\Delta \mu $
in panels (a) and (c) of the figure varies precisely in interval Eq. (\ref%
{3/16}), in agreement with limit relations (\ref{Nlimits}). The figure shows
that there are two threshold values for $N$ and $\Omega $, namely, $N^{%
\mathrm{c1,c2}}$ and $\Omega ^{\mathrm{c1,c2}}$, with $N^{\mathrm{c1}}<N^{%
\mathrm{c2}}$ and $\Omega ^{\mathrm{c1}}<\Omega ^{\mathrm{c2}}$. When $N<N^{%
\mathrm{c1}} $ or $\Omega <\Omega ^{\mathrm{c1}}$, no soliton solutions are
found, only spatially uniform states being possible; note that $N^{\mathrm{c1%
}}=\Omega ^{\mathrm{c1}}=0$ for $S_{+}=0$ [these findings agree with the
known fact that 2D vortex-soliton solutions of Eq. (\ref{stablephi+}) exist
if their norm exceeds a certain threshold value, while there is no existence
threshold for zero-vorticity solutions \cite{Michinel2}. In intervals $N^{%
\mathrm{c1}}<N<N^{\mathrm{c2}}$ and $\Omega ^{\mathrm{c1}}<\Omega <\Omega ^{%
\mathrm{c2}}$, soliton solutions can be found, but they are unstable (as
shown by dashed and dotted curves in Fig. \ref{properties}). Finally, stable
solutions exist at $N>N^{\mathrm{c2}}$ and $\Omega >\Omega ^{\mathrm{c2}}$.
These conclusions for the 2D cubic-quintic system are different from the
case of its counterpart with the cubic-only nonlinearity, where the SV
solutions with $S_{+}=0$, $S_{-}=1$ [see Eq. (\ref{S+S-})] are stable in the
entire bandgap \cite{{Sakaguchi2016}}.

\subsection{Mobility of the gap solitons initiated by kicking}

Construction of moving solitons in the framework of Eqs. (\ref{1}) and (\ref%
{2}) is a nontrivial problem, as they are not Galilean invariant equations.
We have studied the mobility in direct simulations via Eqs. (\ref{1}) and (%
\ref{2}), respectively, by applying kick $\eta $ to the stationary solution
in the $x$ or $y$ direction. According to the relationship in Eq. (\ref%
{J2symmetry}), the corresponding inputs are defined as
\begin{equation}
\psi _{\pm }(x,y,t=0)=%
\begin{cases}
\phi _{\mp }^{\ast }(x,y)\{e^{i\eta x},e^{i\eta y}\} & \mathrm{for\quad Eq.(%
\ref{1}),} \\
\phi _{\pm }(x,y)\{e^{i\eta x},e^{i\eta y}\} & \mathrm{for\quad Eq.(\ref{2}).%
}%
\end{cases}
\label{kicked}
\end{equation}%
The substitution of these expressions in Eq. (\ref{P}) yields the respective
values of the momentum produced by the kick,%
\begin{equation}
P_{x,y}=N\eta .  \label{Pxy}
\end{equation}%
For convenience, we name stable solutions of Eq. (\ref{1}), which are
located near the top edge of the bandgap, as \textquotedblleft top-type
solitons", while the stable solutions of Eq. (\ref{2}), which are located
near the bottom edge of the bandgap, are named as \textquotedblleft
bottom-type solitons".

\begin{figure}[tp]
\centering{\ \includegraphics[scale=0.5]{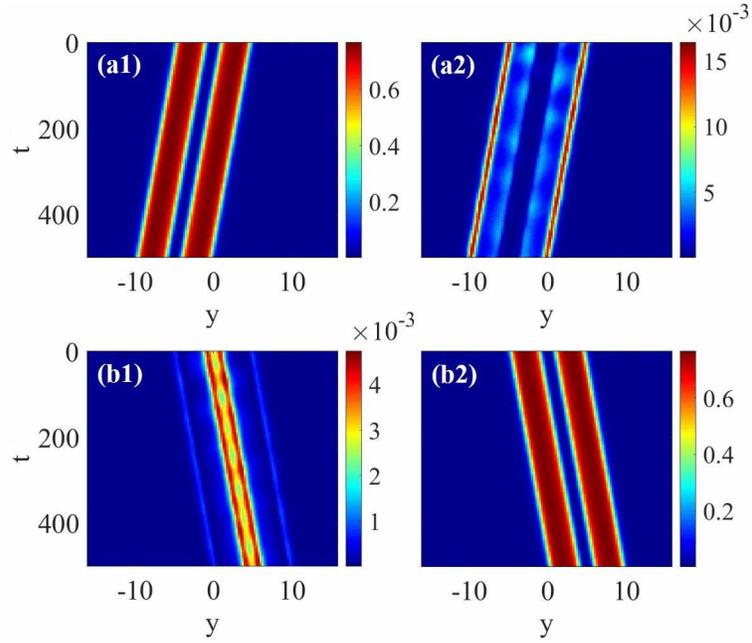}}
\caption{(Color online) Motion of the soliton at the bottom (a1,a2) and top
(b1,b2) of the bandgap initiated by kick $\protect\eta =0.1$ in the positive
$y$ direction, see Eq. (\protect\ref{kicked}). The first row is the motion
of the soliton located near the bottom of the bandgap, with parameters $%
(N_{+},S_{+},\Omega )=(50,1,10)$. The second row is the same for the soliton
located close to the top of the bandgap, with $(N_{-},S_{-},\Omega
)=(50,1,10)$. }
\label{kick1}
\end{figure}

\begin{figure}[tp]
\centering{\ \includegraphics[scale=0.45]{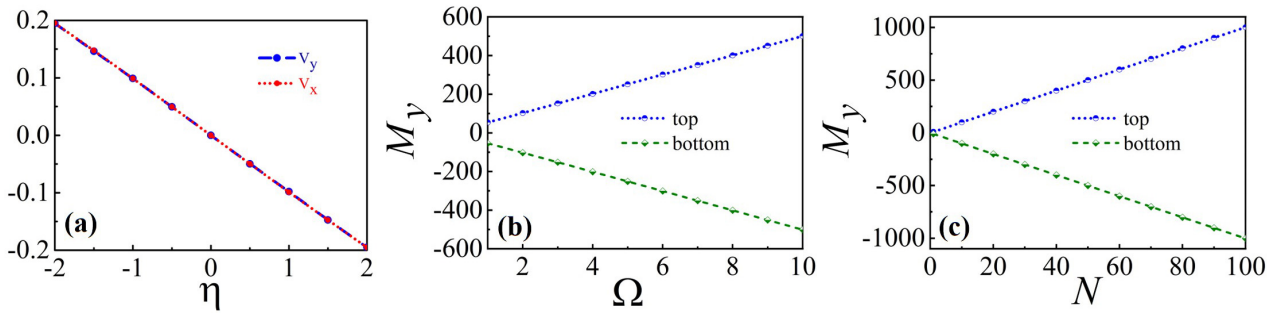}}
\caption{(Color online) (a) The velocity of the gap soliton, which is kicked
along the positive $x$ (red) or $y$ (blue) direction, according to Eq. (%
\protect\ref{kicked}), versus $\protect\eta $. The velocity is found
numerically as per Eqs. (\protect\ref{centermass}) and (\protect\ref%
{Velocity}) The soliton is taken near the bottom of the bandgap, with $%
(N_{+},S_{+},\Omega )=(50,0,10)$. (b) The effective mass of the solitons
located near the top and bottom of the bandgap (short-dashed and dashed
lines, respectively) versus the ZS strength, $\Omega $. Parameters of the
solitons located near the bottom and top edges of the bandgap types are
selected as $(N_{\pm },S_{\pm },\protect\eta )=(50,0,0.1)$. (c) The
effective mass of the solitons located near the top and bottom of the
bandgap (short-dashed and dashed lines, with $(\Omega,S_{\pm },\protect\eta %
)=(10,0,0.1)$, respectively) versus the total norm $N$. The masses are found
as per Eq. (\protect\ref{mass}). }
\label{meff}
\end{figure}

Figure \ref{kick1} displays cross-sections of the two components of the
moving solitons located close to the bottom (a1,a2) and top (b1,b2) of the
bandgap, applying $\eta =0.1$ in the positive $y$ direction, as per Eq. (\ref%
{kicked}). The figure shows that the solitons in these two cases (for the
solitons close to the bottom and top of bandgap) move steadily along the $y$
directions, which indicates that they have a negative and positive effective
mass, respectively. The signs of the effective mass at the two edges of the
bandgap are consistent with previous results \cite{Li2017}.

To accurately identify the effective mass of the solitons from the numerical
data, as the ratio of momenta given by Eq. (\ref{Pxy}) [see Eq. (\ref{P})]
to the corresponding velocities $V_{x,y}$, the center-of-mass (COM)
coordinate of the soliton's dominant component is calculated as
\begin{equation}
\{X_{\mathrm{COM}}(t),Y_{\mathrm{COM}}(t)\}=\frac{\int \int \{x,y\}|\psi _{%
\mathrm{d}}(x,y,t)|^{2}dxdy}{\int \int |\psi _{\mathrm{d}}(x,y,t)|^{2}dxdy}.
\label{centermass}
\end{equation}%
Then, the velocity is found as
\begin{equation}
V_{x}=[X_{\mathrm{COM}}(t)-X_{\mathrm{COM}}(0)]/t,\quad V_{y}=[Y_{\mathrm{COM%
}}(t)-Y_{\mathrm{COM}}(0)]/t.  \label{Velocity}
\end{equation}%
Eventually, effective masses are obtained from here as
\begin{equation}
M_{x,y}=\frac{P_{x,y}}{V_{x,y}}\equiv \frac{N\eta }{V_{x,y}}.  \label{mass}
\end{equation}
The result, $V_{x}=V_{y}$, as displayed in Fig. \ref{meff}(a), implies that,
naturally, the effective mass is isotropic in the present system, i.e., $%
M_{x}=M_{y}$. Figs. \ref{meff}(b,c) show that, as mentioned above, the
solitons located near the bottom and top edges of the bandgap have negative
and positive effective masses, respectively. Moreover, the numerical
results, displayed in these figures, demonstrate that the effective masses
linearly depend on $\Omega $ and $N$, and do not depend on winding numbers
$S_{\pm } $. The increase of the masses with the increase of the ZS
strength, $\Omega $, is explained by the fact that gap solitons do not
exist at $\Omega =0$. The linear dependence of the masses on $N$ is a
common property of solitons in models of the GPE type.

\begin{figure}[tp]
\centering{\ \includegraphics[scale=0.5]{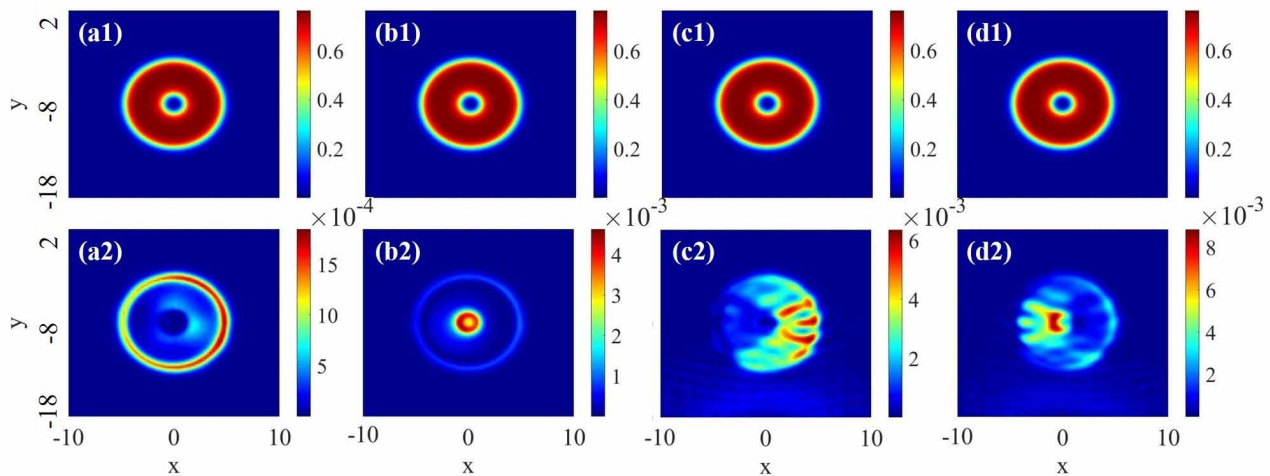}}
\caption{(Color online) The density distribution in the moving solitons
kicked in the positive $y$ direction with different values of $\protect\eta $
as per Eq. (\protect\ref{kicked}). The first and second rows display the
dominant and subordinate components, respectively. Parameters of the soliton
in panels (a1,a2,c1,c2) are $(N_{+},S_{+},\Omega )=(50,1,10)$, while for the
soliton in panels (b1,b2,d1,d2) they are $(N_{+},S_{+},\Omega )=(50,-1,10)$.
The first and second columns display the solitons kicked with $\protect\eta %
=0.1$ and measured at $t=800$, while the third and fourth columns pertain to
the solitons kicked with a kick which is ten times as large, therefore the
data are taken at time which is ten times smaller, $t=80$.}
\label{kick2}
\end{figure}

\begin{figure}[tp]
\centering{\ \includegraphics[scale=0.55]{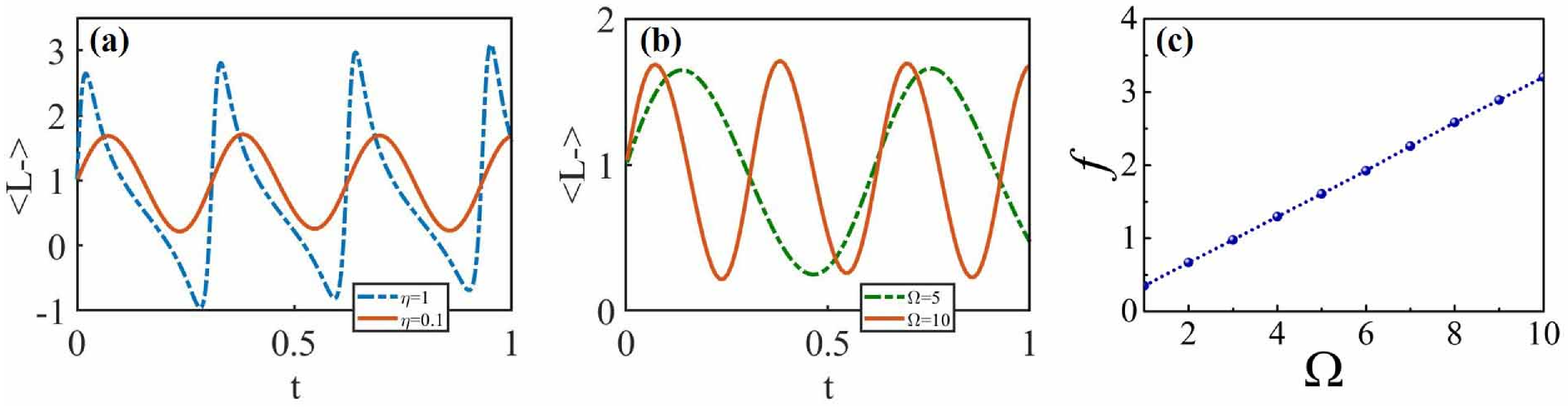}}
\caption{(Color online) (a) NOAM of the subordinate component of the soliton
of the bottom-edge type versus $t$ for different values of the kick, i.e., $%
\protect\eta =0.1$ and $1$ (the red solid and blue dashed curves,
respectively). Parameters of the soliton are $(N_{+},S_{-},\Omega
)=(50,1,10) $. (b) NOAM of a moving soliton with $(N_{+},S_{-},\protect\eta %
)=(50,1,0.1)$ and different values of the ZS strength, i.e., $\Omega =5$ and
$10$ (the green dashed and red solid curves, respectively). (c) The
oscillation frequency of the NOAM of the subordinate component versus $%
\Omega $. Here, we also select the bottom-edge type of the soliton, with
parameters $(N_{+},S_{+},\protect\eta )=(50,0,0.1)$.}
\label{frequency}
\end{figure}

Typical examples of the density distribution in the solitons set in motion
by the kick with different values of $\eta $ are shown, for the solitons
taken near the bottom edge of the bandgap, in Fig. \ref{kick2}. The snapshots
are taken at the moment of time when the COM of the soliton kicked in the $y$
direction is located at $y\approx 8$. It is observed that the dominant
component of the moving solution ($\phi _{+}$, in the present case)
maintains its shape, while the subordinate component, $\phi _{-},$ exhibits
a deformation, as a consequence of the absence of the Galilean invariance in
the system. Accordingly, in the case of a small kick $\eta $, the density
distribution of the subordinate component is still confined within the main
body of the moving soliton, see Figs. \ref{kick2}(a2,b2). In contrast, at
large $\eta $, the subordinate components is lagging behind the main body of
the soliton, which gives rise to its gradual delocalization, see Figs. \ref%
{kick2}(c2,d2).

In the case of small $\eta $, the subordinate component shows an asymmetric
density distribution and rotates around the COM in the course of the direct
simulations of the motion. This fact explains density oscillations in the
cross-section diagram in Figs. \ref{kick1}(a2,b1). To characterize the
motion of the soliton, we define the normalized orbital angular momentum
(NOAM) of the two components as
\begin{equation}
\ \left\langle L_{\pm }\right\rangle =\frac{\int \psi _{\pm }^{\ast }\hat{L}%
\psi _{\pm }d\mathbf{r}}{N_{\pm }},  \label{angular}
\end{equation}%
where $\hat{L}=-i(x\partial _{y}-y\partial _{x})$ is the angular-momentum
operator. The direct simulations show that the NOAM of the dominant
component stays nearly constant in the course of the motion, while the NOAM
of the subordinate one oscillates as a function of $t$. For small $\eta $,
the variation of $\left\langle L_{-}(t)\right\rangle $ manifests itself as
harmonic oscillations [see the red curve in Fig. \ref{frequency}(a)], while
for large $\eta $, the oscillations of $\left\langle L_{-}(t)\right\rangle $
are anharmonic due to the above-mentioned trend to delocalization of the
density distribution, see the blue dashed curve in Fig. \ref{frequency}(a).
Figure \ref{frequency}(b) displays the harmonic oscillations of $%
\left\langle L_{-}(t)\right\rangle $ for different values of $\Omega $.
Further, the numerical results show that the frequency of the harmonic
oscillations linearly depends on the ZS strength $\Omega $ for a fixed value
of the norm of the subordinate component, see Fig. \ref{frequency}(c). On
the other hand, the oscillation frequency does not depend on the winding
number. Because the frequency of the harmonic oscillation is determined by
the oscillator's mass, the conclusion is the same as produced above on the
basis of different numerical data in Fig. \ref{meff}(b): the effective mass
of the soliton linearly grows with $\Omega $, but it does not depend on the
topological charge of the soliton.

\subsection{Collisions between moving gap solitons}

\begin{figure}[th]
\centering
{\includegraphics[scale=0.6]{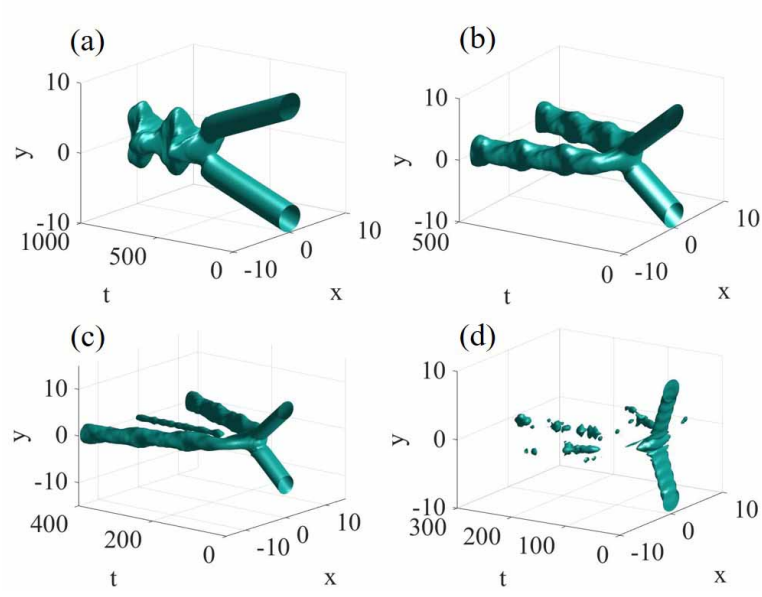}}
\caption{(Color online) Collisions of fundamental solitons (ones with $%
S_{+}=0$). Panels feature the collisions between the solitons set in motion
as per Eq. (\protect\ref{bottomcollideEq}) with (a) $\protect\eta =0.1$, (b)
$\protect\eta =0.5$, (c) $\protect\eta =0.9$, and (d) $\protect\eta =3$.
Here, parameters of the gap solitons are selected as $(N_{+},\Omega
)=(10,10) $.}
\label{collide1}
\end{figure}

\begin{figure}[th]
\centering
{\includegraphics[scale=0.6]{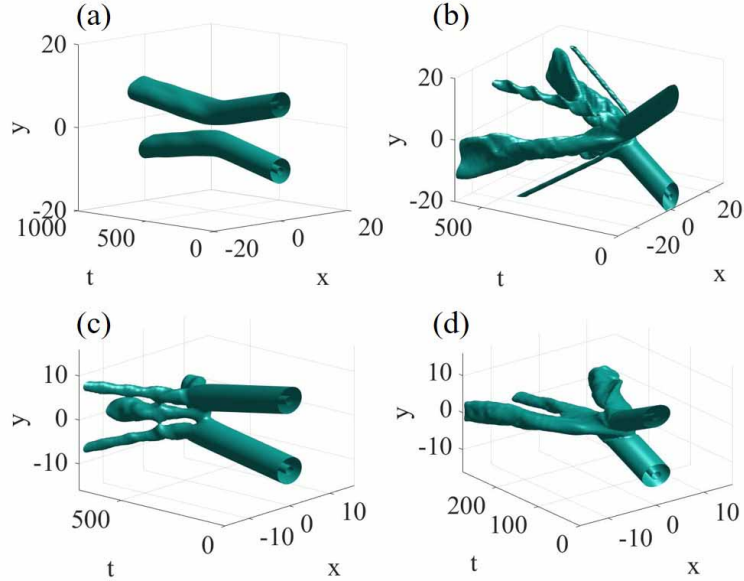}}
\caption{(Color online) Parameters of the colliding FVGSs in this figure are
selected as $(N_{+},\Omega )=(20,10)$. (a,b) The colliding gap solitons have
the same topological charge, with $S_{+}=1$. (a) An elastic collision
between two gap solitons with $\protect\eta =0.1$. (b) An inelastic
collision with $\protect\eta =0.7$. (c,d) Collisions of the solitons with
opposite topological charges, $S_{+}=\pm 1$. The kick corresponding to Eq. (%
\protect\ref{bottomcollideEq}) is $\protect\eta =0.1$ in (c), and $\protect%
\eta =0.7$ in (d).}
\label{collide2}
\end{figure}

A straightforward possibility is to simulate collisions between solitons
which are set in motion by opposite kicks $\pm \eta $. For the solitons
located near the bottom and top edges of the bandgap, we define the
respective initial wave functions as

\begin{equation}
\psi _{\pm }(x,y,t=0)=\phi _{\pm }(x,y+y_{0})e^{-i\eta y}+\phi _{\pm
}(x,y-y_{0})e^{i\eta y},  \label{topcollideEq}
\end{equation}%
and%
\begin{equation}
\psi _{\pm }(x,y,t=0)=\phi _{\pm }(x,y+y_{0})e^{i\eta y}+\phi _{\pm
}(x,y-y_{0})e^{-i\eta y},  \label{bottomcollideEq}
\end{equation}%
where $2y_{0}$ is the initial separation between the solitons. Due to the
symmetry provided by Eq. (\ref{J2symmetry}), we here consider only the
collision between the solitons of the bottom-edge type.

Figures \ref{collide1} and \ref{collide2} display several typical examples
of collisions between the fundamental solitons and FVGSs, with different
values of $\eta $, which are displayed by the contour surface of the density
of the dominant component. In Fig. \ref{collide1}(a), a completely inelastic
collision is observed for a small value of $\eta $, with the colliding
fundamental solitons merging into a breather. On the contrary, in Fig. \ref%
{collide1}(b) a \textquotedblleft twisted" quasi-elastic collision is
observed for a moderate value of $\eta $: the colliding fundamental solitons
first merge and then separate into ones moving perpendicular to the initial
direction. It is further shown in Fig. \ref{collide1}(c) that three
fragments appear after the collision when $\eta $ is large, a new fragment
being a small-amplitude breather. Finally, it is shown in Fig. \ref{collide1}%
(d) that the solitons are completely destroyed by the collision if $\eta $
is sufficiently large.

The collision between two vortex solitons (FVGSs) shows different dynamics.
In Figs. \ref{collide2}(a,b), the colliding solitons have identical
topological charges, $S_{+}=1$. When $\eta $ is small, the solitons repel
each other and maintain their shapes after the rebound, i.e., they collide
elastically. In contrast, in the case of opposite topological charges, the
colliding solitons also repel each other, but the outcome is different: they
split into four fragments, as shown in Fig. \ref{collide2}(c).

When $\eta $ is large, the collisions of the vortex solitons are always
inelastic. In the case of identical topological charges, collision-generated
fragments are hurled in all directions, see Fig. \ref{collide2}(b). In the
case of opposite topological charges, three fragments are generated by the
collision, see Fig. \ref{collide2}(d). If $\eta $ is sufficiently large, the
colliding FVGSs are completely destroyed by the collision. The latter case
is not shown here.

Thus, the present system, based on Eq. (\ref{2}), which is dominated by the
SOC energy demonstrates the behavior completely different from what is
common for usual models with the domination of the kinetic energy:
collisions with large values of velocities tend to be destructive, rather
than quasi-elastic.

\section{Conclusions}

In this work, we have investigated the formation of stable FVGSs
(full-vortex gap solitons) and their dynamics in the two-component BEC with
SOC (spin-orbit coupling) and competing cubic and quintic interactions. The
difference from the previously studied 2D solitons in the SOC system with
the cubic attractive interactions is that they might be stable solely in the
SV (semi-vortex) form, with the vorticity carried by a single component.
When the SOC and ZS (Zeeman-splitting) terms are sufficiently strong, the
kinetic-energy ones may be neglected in the system of coupled GPEs
(Gross-Pitaevskii equations). The so simplified system maintains a bandgap
in its linear spectrum, populated by FVGSs. They may be stable near edges of
the bandgap. If the absolute values of the topological charge of the
dominant component is larger than that of the secondary one, the density
distribution of the secondary component is surrounded by that of the
dominant component. Otherwise, the dominant component is surrounded by the
secondary one. We represent FVGS families by displaying dependences of the
chemical potential and solitons' effective area on the norm and ZS strength.
The results reveal stability boundaries for the families. The mobility of
the FVGSs is explored by kicking the stationary solitons. It is thus found
that the FVGSs formed near the top and bottom edges feature, respectively,
positive and negative dynamical masses. For a small kick, the minor
component of solitons oscillates with the frequency proportional to the ZS
strength. In contrast, under the action of a larger kick the minor suffers
strong deformation, eventually leading to delocalization of the kicked
soliton. Collisions between FVGSs with identical and opposite topological
charges were considered too, featuring both elastic and inelastic outcomes.

A natural direction to extend the present work is to investigate
self-trapped modes of the FVGSs type in a system with long-range
dipole-dipole interactions, as well as in three-dimensional systems.

\begin{acknowledgments}
This work was supported by NNSFC (China) through grant Nos. 11905032 \&
11874112, Natural Science Foundation of Guangdong province through grant
No.2021A1515010214, the GuangDong Basic and Applied Basic Research
Foundation through grant No. 2021A1515111015, the Key Research Projects of
General Colleges in Guangdong Province through grant No. 2019KZDXM001, the
Foundation for Distinguished Young Talents in Higher Education of Guangdong
through grants Nos. 2018KQNCX279 \& 2018KTSCX241, the Special Funds for the
Cultivation of Guangdong College Students Scientific and Technological
Innovation through grant Nos. pdjh2021b0529 \& pdjh2022a0538, the Research
Fund of Guangdong-Hong Kong-Macao Joint Laboratory for Intelligent
Micro-Nano Optoelectronic Technology through grant No.2020B1212030010 and
the Graduate Innovative Talents Training Program of Foshan University. The
work of B.A.M. is supported, in part, by the Israel Science Foundation
through grants No. 1286/17 and 1695/22.
\end{acknowledgments}

\end{document}